\documentclass[twocolumn,superscriptaddress,longbibliography,
	aps,preprintnumbers]{revtex4-1}

\usepackage{graphicx}
\usepackage{bm}
\usepackage{color}
\usepackage{epstopdf}
\usepackage{amsmath}
\usepackage{amssymb}
\usepackage{epstopdf}

\usepackage[urlcolor=blue,colorlinks=true,citecolor=blue,
	linkcolor=blue,pdfstartview={FitH},bookmarks=false]{hyperref}

\graphicspath{{fig/}{./fig/}{.}}

\begin{document}

\setcounter{figure}{0}
\renewcommand{\thefigure}{\arabic{figure}}

\title{Delocalisation of Majorana quasiparticles in plaquette--nanowire hybrid system}

\author{Aksel Kobia\l{}ka}
\email[e-mail: ]{akob@kft.umcs.lublin.pl}
\affiliation{Institute of Physics, M. Curie-Sk\l{}odowska University, \\ 
pl. M. Sk\l{}odowskiej-Curie 1, PL-20031 Lublin, Poland}

\author{Tadeusz Doma\'{n}ski}
\email[e-mail: ]{doman@kft.umcs.lublin.pl}
\affiliation{Institute of Physics, M. Curie-Sk\l{}odowska University, \\ 
pl. M. Sk\l{}odowskiej-Curie 1, PL-20031 Lublin, Poland}

\author{Andrzej Ptok}
\email[e-mail: ]{aptok@mmj.pl}
\affiliation{Institute of Nuclear Physics, Polish Academy of Sciences, \\ 
ul. E. Radzikowskiego 152, PL-31342 Krak\'{o}w, Poland}

\date{\today}

\begin{abstract}
Interplay between superconductivity, spin-orbit coupling and magnetic field can lead to realisation 
of the topologically non--trivial states which in finite one dimensional nanowires are 
manifested by emergence of a pair of zero-energy Majorana bound states.
On the other hand, in two dimensional systems the chiral edge states can appear. 
We investigate novel properties of the bound states in a system of {\it mixed 
dimensionality}, composed of one-dimensional nanowire connected with two-dimensional plaquette.
This setup could be patterned epitaxially, e.g.\ using heterostructure analogous to what has been 
reported recently by F. Nichele {\it et al.}, \href{http://doi.org/10.1103/PhysRevLett.119.136803} 
{Phys. Rev. Lett. {\bf 119},~136803~(2017)}. 
We study this system, assuming either its part or the entire 
structure to be in topologically non--trivial superconducting state. 
Our results predict delocalisation of the Majorana modes, upon leaking from the nanowire to the nanocluster with some tendency  towards its corners.
\end{abstract}

\maketitle

\section{Introduction}

Recent nanotechnological progress  allows for fabrication of  artificial nanostructures~\cite{santos.deen.15}, 
where unique quantum phenomena and new states of matter~\cite{soumyanarayanan.reyren.16} could be observed.
Prominent examples are the Majorana bound states (MBS) of quasi-one-dimensional structures, 
e.g.\ semiconducting--superconducting hybrid nanowire~\cite{deng.yu.12,mourik.zuo.12,das.ronen.12,finck.vanharlingen.13,deng.vaitiekenas.16,nichele.drachmann.17,lutchyn.bakkers.18,gul.zhang.18} or nanochains of magnetic atoms  deposited on superconducting surface~\cite{nadjperge.drozdov.14,pawlak.kisiel.16,feldman.randeria.16,ruby.heinrich.17,jeon.xie.17,kim.palaciomorales.18}.
Such MBS are characterized by  particle--antiparticle indistinguishability and their non--Abelian statistics~\cite{nayak.simon.08} makes them promising for realisation of the topological quantum computing~\cite{aasen.hell.16,karzig.knapp.17}.

Proximity-induced superconductivity combined with the magnetic field and the spin-orbit 
coupling (SOC) drives the system from its topologically trivial to non--trivial superconducting 
phase~\cite{klinovaja.loss.12}. Such transition occurs at  critical magnetic field $h_{c}$, 
dependent on the SOC strength and dimensionality of the system~\cite{sato.fujimoto.09,sato.takahashi.09,sato.takahashi.10}.
Spectroscopically it is manifested  by a coalescence of one pair of the Andreev (finite-energy) bound 
states into the Majorana (zero--energy) quasiparticles~\cite{chevallier.sticlet.12,chevallier.simon.13}.

Emergence of the degenerate Majorana modes from the Andreev bound states has been also reported in hybrid structures, comprising the quantum dots side-attached to the topological superconducting nanowires~\cite{deng.vaitiekenas.16}.
Initial theoretical prediction of such MBS {\em leakage} on the quantum dot region~\cite{vernek.penteado.14}
has been investigated  by various groups~\cite{liu.sau.17,hoffman.chevallier.17,ptok.kobialka.17,chevallier.szumniak.18,stenger.woods.18,kobialka.ptok.18b,moore.stanescu.18,gorski.baranski.17,fleckenstein.dominguez.18}.
In these hybrid structures the wavefunction of Majorana quasiparticle is spread onto a region 
of the normal quantum dot--superconducting nanowire interface~\cite{gibertini.taddei.12,klinovaja.loss.12,chevallier.sticlet.12,guigou.sadlmayr.16,kobialka.ptok.18,fleckenstein.dominguez.18},
diluting the spatial distribution of its spectral weight. 
This issue has been a subject of intensive experimental and theoretical studies~\cite{deng.vaitiekenas.18}.

\begin{figure}[!b]
\includegraphics[width=0.75\linewidth]{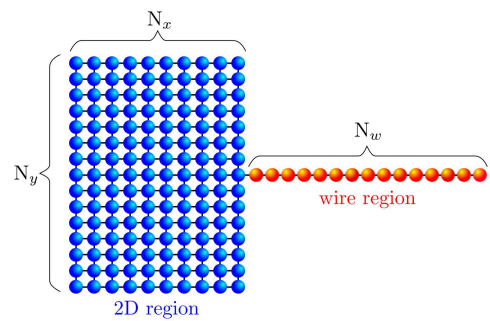}
\caption{
\label{fig.schem}
Scheme of the hybrid structure, where a semiconducting Rashba nanowire (comprising $N_{w}$ sites) 
is connected to 2D cluster (whose dimensions are $N_{x} \times N_{y}$). This system is deposited 
on a surface of the superconducting substrate.}
\end{figure}

\begin{figure*}
\centering
\includegraphics[width=0.7\linewidth]{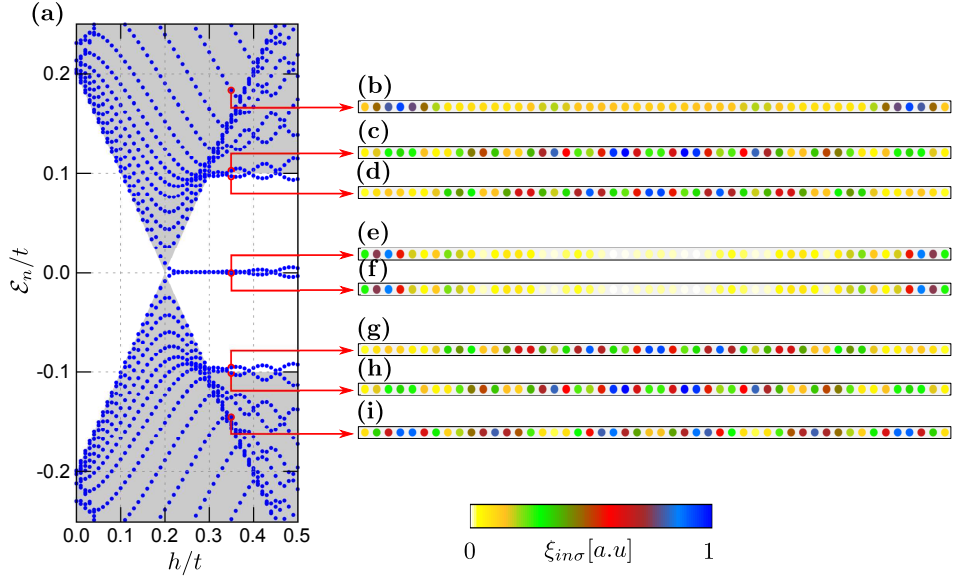}
\caption{
\label{fig.1d}
Low energy spectrum of the 1D nanowire (a) and the spatially resolved probabilities 
$\xi_{in}=\sum_{\sigma}\xi_{in\sigma}$ for the quasiparticle energies $\mathcal{E}_{n}$ indicated 
by the red arrows (b)-(i). Results are obtained for $N_{x} = N_{y} = 0$, $N_{w} = 50$ and $\mu_{w} = -2.0t$.
}
\end{figure*}

\begin{figure}
\centering
\includegraphics[width=\linewidth]{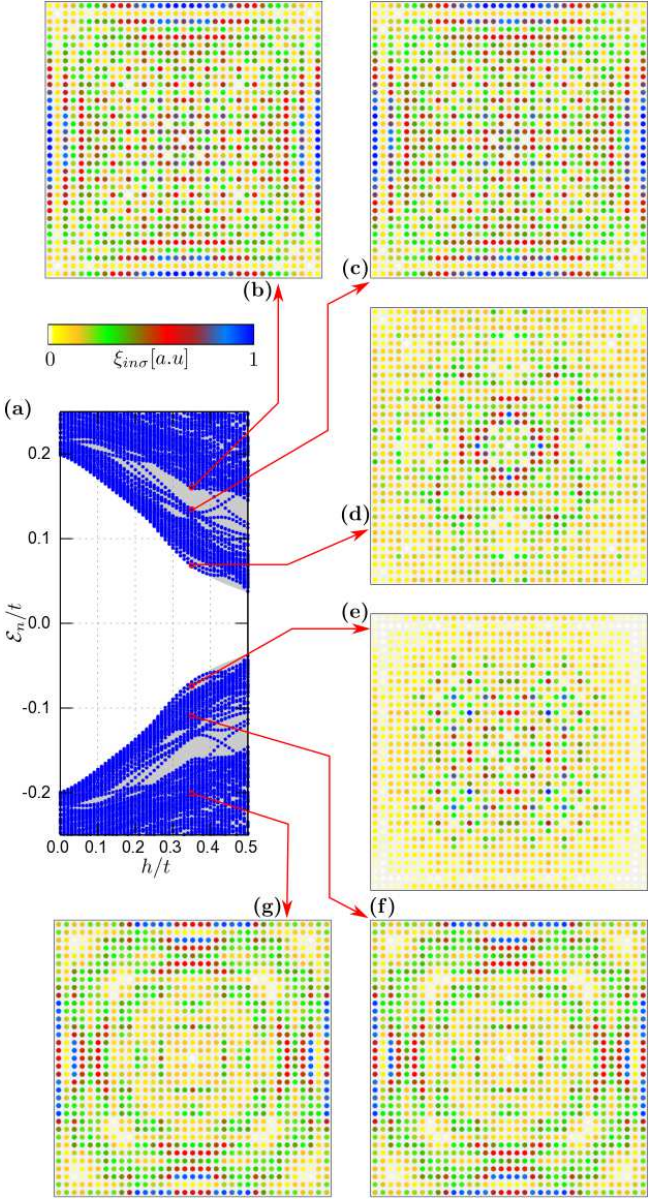}
\caption{
\label{fig.2d.triv}
Low energy spectrum of the 2D plaquette obtained in topologically trivial state 
for $\forall_{i} \mu_{i} = -2.0 t$ (a). Panels (b)-(g) display spatial profiles 
of the quasiparticle for several eigenvalues $\mathcal{E}_{n}$. Results are 
obtained for $h / t = 0.35$, assuming $N_{x} = N_{y} = 35$ and $N_{w} = 0$.}
\end{figure}

\begin{figure}
\centering
\includegraphics[width=\linewidth]{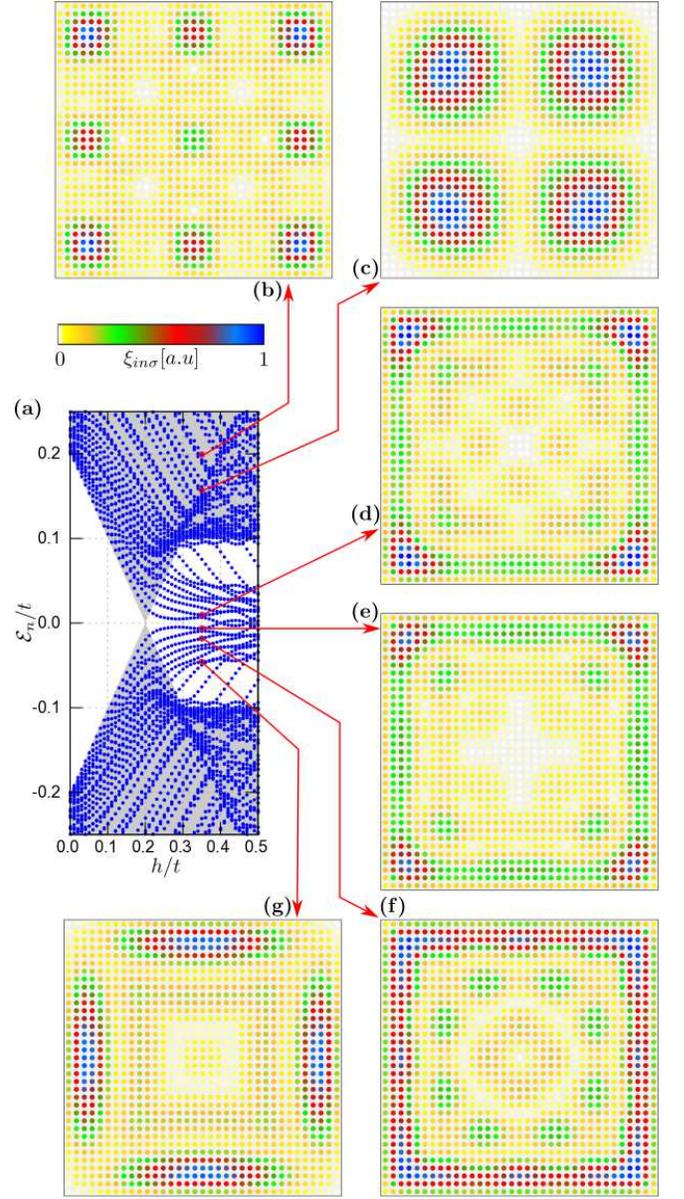}
\caption{
\label{fig.2d}
Low energy spectrum of the 2D plaquette obtained in topologically non-trivial 
state for  $\forall_{i} \mu_{i} = -4.0 t$ (a). All other model parameters are 
the same as in Fig.~\ref{fig.2d.triv}.}
\end{figure}

\begin{figure*}
\centering
\includegraphics[width=\linewidth]{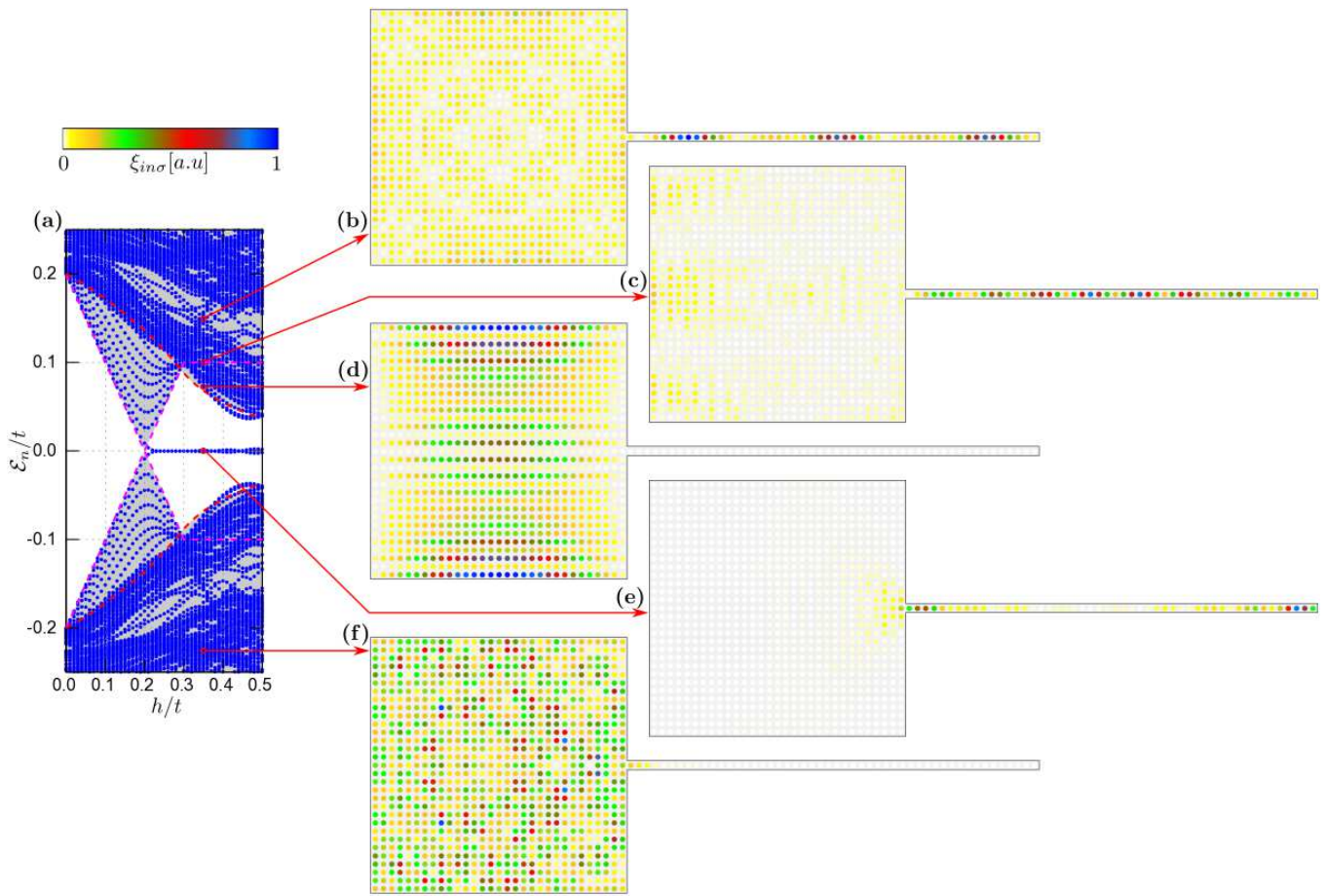}
\caption{
\label{fig.sys1}
Spectrum of the plaquette--nanowire hybrid system obtained for $N_{x} = N_{y} = 31$ 
and $N_{w} = 50$, $h = 0.35 t$, assuming $\mu_{2D} = \mu_{w} = -2.0 t$. In this case, 
only the nanowire is in the non-trivial topological superconducting phase. 
}
\end{figure*}

\begin{figure*}
\centering
\includegraphics[width=\linewidth]{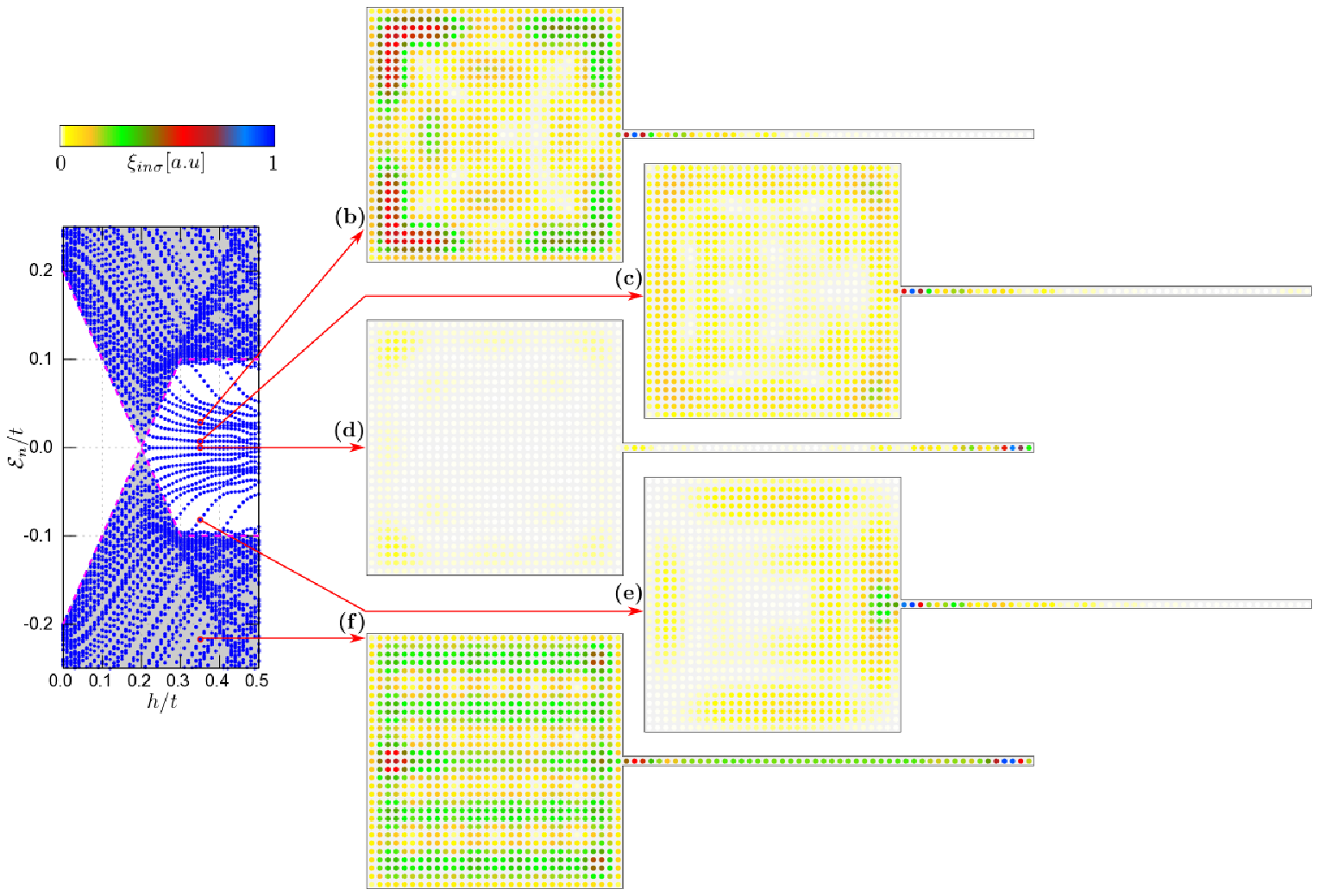}
\caption{
\label{fig.sys2}
Spectrum of the plaquette--nanowire hybrid system obtained for $N_{x} = N_{y} = 31$ 
and $N_{w} = 50$, $h = 0.35 t$, assuming $\mu_{2D} = -4.0 t$ and $\mu_{w} = -2.0 t$, 
which guarantee that both constituents are in the non--trivial topological phase.
}
\end{figure*}

In one-dimensional structures the Majorana quasiparticles localise at the sample
boundaries~\cite{kitaev.01} or near internal defects \cite{maska_etal.17,kobialka.ptok.18b,ptok.cichy.18}. Contrary 
to that, for quasi two-dimensional systems there have been predicted  chiral edge 
modes~\cite{rontynen.15,bjornson.15,li.16,rachel.mascot.17} enabling the Majoranas to be 
delocalised, both in the real and momentum spaces~\cite{kobialka.ptok.18b}. 
Evidence for such dispersive Majorana modes have been recently provided by 
STM measurements for magnetic islands deposited on superconducting substrates
\cite{menard.17,He_etal.2017,wiesendanger_etal.18,PascalSimon_etal.18}. Another 
route to achieve the topological superconductivity and MBS in two-dimensional 
systems relies on the phase biased planar Josephson junctions, which confine 
the narrow strip of electron gas subject to the Rashba interaction and magnetic field  
\cite{fornieri_etal.18,ren_etal.18}. 

In general,  realisation of the MBS might not be restricted solely to systems with 
simple geometries~\cite{bjornson.blackschaffer.16,stanescu.dassarma.18,morr.17} therefore we 
propose the setup of {\it mixed dimensionality}, comprising one-dimensional nanowire 
coupled to  two-dimensional plaquette (Fig.~\ref{fig.schem}). This situation resembles 
the recently investigated nanostructures, where quasi-one-dimensional wires are attached 
to larger structures~\cite{nichele.drachmann.17}. We study here subgap spectrum of this 
system, focusing on spatial profiles of the Majorana modes leaking from the nanowire 
into the adjoined plaquette. We explore the quasiparticle spectra of this setup for 
representative values of the chemical potentials of both constituents (tunable by 
electrostatic potentials), which control topological nature of their superconducting 
phase. Our study gives an insight into non--local character of the Majorana 
quasiparticles.

The paper is organised as follows. In Sec.~\ref{sec.model} we introduce the microscopic
model and present  computational details. In Sec.~\ref{sec.num_res}, we describe 
the numerical results obtained for each constituent  and for the entire hybrid structure. 
Beyond the scope of numerical calculations, we also described the proposal for experimental verification of out theoretical predictions (Sec.~\ref{sec.proposal}).
Finally, we summarise the results in Sec.~\ref{sec.sum}.

\section{Model and method}
\label{sec.model}

The nanostructure shown in Fig.~\ref{fig.schem} can be modelled by the real space Hamiltonian 
$\mathcal{H} = \mathcal{H}_{kin} + \mathcal{H}_{sc} + \mathcal{H}_{soc}$.
The first term describes the kinetic energy:
\begin{eqnarray}
\mathcal{H}_{kin} = \sum_{ij\sigma} \left\lbrace - t \delta_{\langle i,j \rangle} 
- \left( \mu_{i} + \sigma h \right) \delta_{ij} \right\rbrace c_{i\sigma}^{\dagger} c_{j\sigma},
\end{eqnarray}
where $t$ denotes the hopping integral between nearest-neighbour sites and
$c_{i\sigma}^{\dagger}$ ($c_{i\sigma}$) describes creation (annihilation) of
electron on {\it i}-th site with spin $\sigma$. In general, the chemical 
potential $\mu_{i}$ can be tuned \textit{in-situ} by some external gate voltage. 
For simplicity, however, we assume it to be constant over the entire 2D plaquette 
($\forall_{i \in 2D} \; \mu_{i} = \mu_{2d}$) and in the 1D nanowire 
($\forall_{i \in w} \; \mu_{i} = \mu_{w}$). We assume the magnetic field $h$ 
to be parallel along the wire and neglect any orbital effects~\cite{kiczek.ptok.17}.
The second term
\begin{eqnarray}
\mathcal{H}_{sc} = \sum_{i} \Delta ( c_{i\downarrow} c_{i\uparrow} +   
c_{i\uparrow}^{\dagger} c_{i\downarrow}^{\dagger} ),
\end{eqnarray}
accounts for the proximity induced on-site pairing, where $\Delta$ is 
the uniform energy gap in the system. The spin--orbit coupling (SOC) 
is given by~\cite{li.covaci.11,li.covaci.12,smith.tanaka.16,goertzen.tanaka.17}
\begin{eqnarray}
\nonumber 
\mathcal{H}_{soc} = \lambda \sum_{i \sigma\sigma'} ( i c_{i\sigma}^{\dagger} 
\sigma_{x}^{\sigma\sigma'} c_{i+\hat{y},\sigma'} - i c_{i\sigma}^{\dagger} 
\sigma_{y}^{\sigma\sigma'} c_{i+\hat{x},\sigma'} + \mbox{\rm H.c.}), \\
\end{eqnarray}
where $\sigma_{i}$ are the Pauli matrices and $\lambda$ stands for the Rashba potential.
Since the nanowire is oriented along $\hat{x}$ axis, only the second part of SOC term survives.

Hamiltonian $\mathcal{H}$ of the hybrid structure can be diagonalised by the Bogoliubov--Valatin 
transformation~\cite{degennes.89}
\begin{eqnarray}
c_{i\sigma} &=& \sum_{n} \left( u_{in\sigma} \gamma_{n} 
- \sigma v_{in\sigma}^{\ast} \gamma_{n}^{\dagger} \right) ,
\label{BdG_transformation}
\end{eqnarray}
where $\gamma_{n}$, $\gamma_{n}^{\dagger}$ are the new {\it quasi}-particle fermionic 
operators and $u_{in\sigma}$, $v_{in\sigma}$ are the corresponding eigenvectors. From 
this transformation (\ref{BdG_transformation}) we get the Bogoliubov--de~Gennes (BdG) 
equations 
\begin{eqnarray}
&& \mathcal{E}_{n}
\left(
\begin{array}{c}
u_{in\uparrow} \\ 
v_{in\downarrow} \\ 
u_{in\downarrow} \\ 
v_{in\uparrow}
\end{array} 
\right) =\\
\nonumber &=& \sum_{j} \left(
\begin{array}{cccc}
H_{ij\uparrow} & D_{ij} & S_{ij}^{\uparrow\downarrow} & 0 \\ 
D_{ij}^{\ast} & -H_{ij\downarrow}^{\ast} & 0 & S_{ij}^{\downarrow\uparrow} \\ 
S_{ij}^{\downarrow\uparrow} & 0 & H_{ij\downarrow} & D_{ij} \\ 
0 & S_{ij}^{\uparrow\downarrow} & D_{ij}^{\ast} & -H_{ij\uparrow}^{\ast}
\end{array} 
\right)
\left(
\begin{array}{c}
u_{jn\uparrow} \\ 
v_{jn\downarrow} \\ 
u_{jn\downarrow} \\ 
v_{jn\uparrow}
\end{array} 
\right) ,
\label{BdG_eqns}
\end{eqnarray}
where $H_{ij\sigma} = - t \delta_{\langle i,j \rangle} - ( \mu_{i} + \sigma h ) \delta_{ij}$ is the single-particle term, 
$D_{ij} = \Delta \delta_{ij}$ refers to the superconducting gap, 
and $S_{ij}^{\sigma\sigma'} = - i \lambda ( \sigma_{y} )_{\sigma\sigma'} \delta_{\langle i,j \rangle}$ is the SOC term (which mixes particles with different spins), where $S_{ij}^{\downarrow\uparrow} = ( S_{ji}^{\uparrow\downarrow} )^{\ast}$ and $S_{ji}^{\uparrow\uparrow} = S_{ij}^{\downarrow\downarrow} = 0$.

From numerical solution of the BdG equations (\ref{BdG_eqns}) we determine the Green's function
$\langle \langle c_{i\sigma} | c_{i\sigma}^{\dagger} \rangle \rangle$ and compute the local 
density of states (LDOS) defined as $\rho_{i,\sigma}( \omega ) = - \frac{1}{\pi} \mbox{Im} \langle 
\langle c_{i\sigma} | c_{i\sigma}^{\dagger} \rangle \rangle$. In the present case we have
\begin{eqnarray}
\rho_{i,\sigma}( \omega ) = \sum_{n} \xi_{in\sigma} \left[ \delta \left( \omega - \mathcal{E}_{n} \right) 
+ \delta \left( \omega + \mathcal{E}_{n} \right) \right] ,
\label{eq.ldos}
\end{eqnarray}
where the spectral weights 
\begin{eqnarray}
\xi_{in\sigma} = | u_{in\sigma} |^{2} \theta ( - \mathcal{E}_{n} ) + | v_{in\sigma} |^{2} 
\theta ( \mathcal{E}_{n} ) 
\end{eqnarray}
refer to probability of the {\it n}-th quasiparticle energy and spin $\sigma$ to exist 
at {\it i}-th site of the system~\cite{glodzik.ptok.18}.

\section{Numerical results}
\label{sec.num_res}

In what follows we study the hybrid consisting of $N = N_{x} \times N_{y} + N_{w}$ 
sites, comprising $N_{x} \times N_{y}$ sites of the 2D--plaquette  and $N_{w}$ sites 
in of 1D--nanowire. For numerical computations we choose $\Delta/t = 0.2$ and 
$\lambda/t=0.15$. In presence of the spin-orbit coupling and the Zeeman effect 
the on-site electron pairing evolves into the inter-site ({\it p}--wave) superconducting 
phase~\cite{sato.takahashi.10,zhang.tewari.08,seo.han.12,ptok.rodriguez.18}. Its topological
form occurs above the critical magnetic field $h_{c} = \sqrt{ \Delta^{2} + ( - W \pm \mu 
)^2 }$~\cite{sato.fujimoto.09,sato.takahashi.09,sato.takahashi.10}, where $W$ is half of 
the bandwidth (equal to $2t$ and $4t$ for 1D and 2D system, respectively). Upon increasing 
the magnetic field $h$, the superconducting gap closes at $h_{c}$ and reopens when entering 
the topological region. Our calculations are done for the finite-size system,
therefore the quasiparticle spectra are discretised. As a useful guide-to-eye, in panels 
(a) of Figs \ref{fig.1d}--\ref{fig.sys2} we have marked the continuous spectrum 
of the bulk system by grey colour.

\subsection{Separate components of the system}

Here, we briefly describe the quasiparticles for each component of
the hybrid structure separately. Let us start with the 1D chain. For 
$\forall_{i} \mu_{w} = -2.0 t$ and assuming the magnetic field $h = 0.35 t$ 
the nanowire would be in its non--trivial topological state. Figs~\ref{fig.1d}(b)--(i) 
show the spatial distribution for representative quasiparticle states whose 
energies are indicted by the arrows. The quasiparticles from outside the topological 
gap [panels (b)--(d) and (g)--(i)] are spread all over the nanowire, whereas two 
states residing inside the topological gap [panels (e) and (f) in Fig.~\ref{fig.1d}] 
are clearly localised near the nanowire ends. Such zero-energy quasiparticles exist 
only in the topologically region, above the critical magnetic field $h_{c} = 0.2t$,
and can be identified as the MBS. Let us notice, that quasiparticles at energies 
$\pm \mathcal{E}_{n}$  have the same spatial patterns [cf.\ panels (e) and (f), 
(d) and (g), or (c) and (h)], due to electron--hole symmetry of the BdG equations.

Let's now focus on properties of the  plaquette. Fig.~\ref{fig.2d.triv} 
shows the spectrum and displays the profiles of selected quasiparticles 
obtained for $\forall_{i} \mu_{i} = -2.0 t$ when the 2D-region is in a trivial 
superconducting phase. Under such circumstances, there is no evidence for 
any in-gap quasiparticles regardless of $h$.  Fig.~\ref{fig.2d} 
presents the results obtained $\forall_{i} \mu_{i} = -4.0 t$, corresponding
to the non-trivial superconducting phase. By inspecting the quasiparticles
outside the topological gap [panels (b)--(c)] we observe their nearly uniform 
distribution in the plaquette without clear signatures of any edge phenomena.
On the other hand, the quasiparticle states existing inside the topological 
gap [panels (d)--(g)] reveal a tendency towards their localisation near 
the sample boundaries~\cite{rontynen.15,bjornson.15,li.16,rachel.mascot.17}.

Appearance of the MBS in the nanowire can be associated with changes of 
the topological $\mathbb{Z}_{2}$ index~\cite{kane.mele.05}. Furthermore, 
we know that the MBS must always appear in pairs in the nanowires. Contrary 
to this, in the 2D-plaquette we observe plenty of in--gap states whose 
energies differ from zero. Strictly speaking, we do not observe 
the completely localised Majorana states in such 2D systems. This 
situation changes qualitatively, however, in the plaquette--nanowire 
hybrid system.

\subsection{Plaquette-nanowire hybrid}

Let us first consider the case, when only part of our hybrid setup is in 
the topologically non--trivial superconducting phase. This can be realised e.g.\ 
for the chemical potential  $\mu_{2D} = \mu_{w} = -2 t$ (Fig.~\ref{fig.sys1}).
For the chosen model parameters, the critical magnetic field of the nanowire 
is $h_{c} = 0.2 t$. In this case the quasiparticle spectrum shows a collection 
of levels, originating from the 1D and 2D regions [cf.\ with Fig.~\ref{fig.1d} 
and~\ref{fig.2d}]. By increasing the magnetic field the original gap closes at 
$h = h_{c}$ and at stronger magnetic fields only the nanowire part is in the 
non--trivial topological phase. In consequence, we observe emergence of just 
one pair of the nearly--zero--energy bound states originating from the 1D part 
of our setup [Fig.~\ref{fig.sys1}(e)]. One of these quasiparticles is localised 
at interface with the nanowire region and partly leaks into the 2D plaquette. 
Other (finite-energy) quasiparticles are distributed either over the plaquette
[Fig.~\ref{fig.sys1}(d)] or over both regions of the setup [Fig.~\ref{fig.sys1}(b) 
and (c)].

We now describe the results for the system, in which both 1D and 2D regions are 
in non--trivial topological phase. This situation can be achieved by fine-tuning 
the site-dependent chemical potentials. For instance $\mu_{w} = -2 t$ and 
$\mu_{2D} = - 4 t$ yield a transition to topological phase of the entire system 
at $h_{c} = 0.2 t$. Numerical results for this case are shown in Fig.~\ref{fig.sys2}. 
By inspecting the lowest in-gap states for $h > h_{c}$ we observe their localisation 
near the boundaries of the system, i.e.\ at a free-standing end of the nanowire 
(right-hand side of the system) and in corners of the plaquette. On the other hand, 
the states from the electron-band regions are nearly uniformly spread over 
the whole structure [Fig.~\ref{fig.sys2}(f)].

\begin{figure}[!t]
\centering
\includegraphics[width=\linewidth]{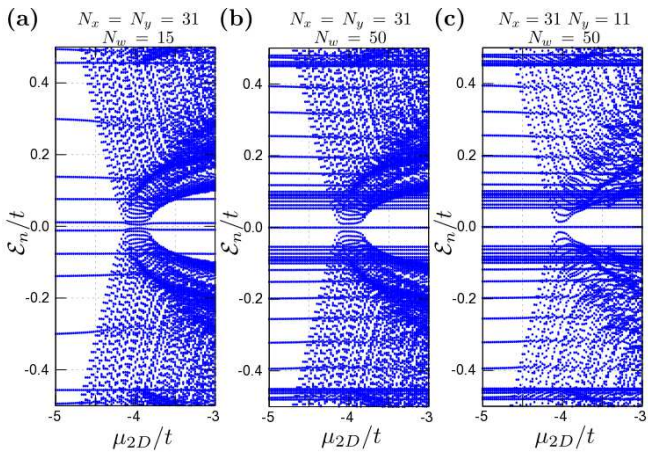}
\caption{
\label{fig.size}
Spectrum of the plaquette--nanowire hybrid as a function of $\mu_{2D}$ for various sizes 
of 1D and 2D components (as indicated) obtained for $h/t = 0.25$ and $\mu_{w}/t = -2$.}
\end{figure}

It is worth noting that both parts of the system have comparable topological 
gap. Due to existence of the common topological non-trivial state, all the 
in-gap states tend to be localised at the sample edges. The quasiparticle 
state appearing at zero energy is predominantly localised in the right hand 
end of the wire  [Fig.~\ref{fig.sys2}(d)], whereas its co-partner (initially 
localised at the left side of nanowire) partly {\it leaks} onto the adjoined
2D-region and appears predominantly in the corners of the plaquette. 
Contrary to the previous case [displayed in Fig.~\ref{fig.sys1}(e)] 
the MBS are strongly delocalised and redistributed. The other finite--energy 
states appear either near the wire-plaquette boundary [see Fig.~\ref{fig.sys2}(c) 
and (e)] or at the edges of the plaquette.

\begin{figure}[!t]
\centering
\includegraphics[width=\linewidth]{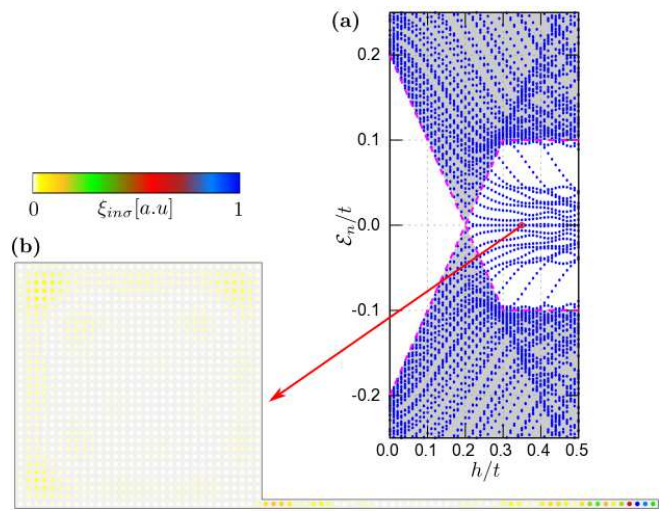}
\caption{
\label{fig.wire_corner}
Spectrum of the plaquette--nanowire hybrid obtained for $N_{x} = N_{y} = 31$, 
$N_{w} = 50$, in the case when the nanowire is connected to the corner of 
the 2D-region.
}
\end{figure}

\paragraph*{Role of finite size effects.} --
Here, we address influence of the finite-size of our hybrid structure.
In Fig.~\ref{fig.size} we show the eigenvalues for three different sizes of the system as a function of the chemical potential $\mu_{2D}$.
Let us remark, that plaquette is in the non--trivial topological phase for $\mu_{2D}/t ~ -4.0$. 
Emergence of the in--gap states is well visible in all cases as can be seen by the horizontal zero-energy lines that correspond to the quasiparticles originating from the nanowire (for which we have fixed the chemical at $\mu_{w}/t = -2$).
If a nanowire is very short, the MBS overlap with each other, forming the bonding and anti-bonding states [Fig.~\ref{fig.size}(a)]. 
Consequence of such overlapping wavefunctions have been studied by a number of authors~\cite{benshach.haim.15,escribano.levyyeyati.17,penaranda.aguado.18}. 
In some analogy to this behaviour, also variation of the plaquette size $N_{x} \times N_{y}$ can lead to rearrangement of the quasiparticle states, depending on $\mu_{2D}$. 
In particular, it may reduce a number of the in--gap states which appear nearby the sample edges.

\paragraph*{Nanowire coupled to plaquette's corner} --
We have checked numerically, that spatial patterns of the MBS leaking from 
1D to 2D parts do not much depend on a particular location of the contact point 
between these constituents. In Fig.~\ref{fig.wire_corner} we illustrate this 
effect, considering the hybrid structure where the nanowire is attached to a corner of the plaquette. 
In this situation the delocalised MBS accumulates near three other corners of the plaquette, yet some of its remnants are still observable at the interface with nanowire.
Irrespectively of the particular contact point the MBS is again strongly delocalised [cf. Fig.~\ref{fig.sys2}(d)].

\section{Proposal for empirical detection}
\label{sec.proposal}

Quasiparticles of the topologically non--trivial superconducting state 
appearing at zero energy at boundaries of one and two-dimensional parts 
in our hybrid structure could be experimentally probed by the scanning 
tunnelling microscopy (STM). Its low-energy and spin-polarised version, 
relying on the {\it selective equal--spin Andreev reflection} (SESAR),  has 
been proposed~\cite{He-2014} as a unique tool for probing the Majorana 
quasiparticles manifested by the zero-bias tunnelling conductance. 
This kind of STM measurements, using ferromagnetic tip, have been 
already done by A.\ Yazdani and coworkers \cite{Yazdani-2017} 
revealing magnetic polarisation of the Majorana modes localised 
near the ends of Fe atoms nanochain  deposited on superconducting Pb. 

Let us briefly explain how such SESAR spectroscopy could probe the spatial 
distribution of the localised and delocalised Majorana quasiparticles 
in our hybrid structure. By applying the voltage $V$ between the conducting 
STM tip and the superconducting nanowire-plaquette system there would be 
induced the charge transport of a given spin $\sigma$ carriers.
On microscopic level,  electrons arriving from the STM tip would
be converted into the inter-site pairs, reflecting holes of the same 
spin polarisation back to the tip. The resulting current can be 
expressed by the Landauer--B\"uttiker formula~\cite{MaskaDomanski-2017}:
\begin{eqnarray} 
\nonumber I_{i,i+1}^{\sigma}(V) = \frac{e}{h} \int  T_{i,i+1}^{\sigma}(\omega)
\left[ f(\omega-eV) - f(\omega+eV)\right] d\omega, \\
\label{I_A}
\end{eqnarray} 
where the transmittance  for a given pair of the neighbouring sites is
$T_{i,i+1}^{\sigma}(\omega) =  \left| \Gamma^{\sigma}_{N} \right|^{2} \;  
\left| \langle\langle \hat{d}_{i\sigma};\, \hat{d}_{{i+1}\sigma}
\rangle\rangle \right|^{2}$
with $\Gamma_{N}^{\sigma}$ denoting the spin-dependent hybridisation between 
the STM tip and individual sites of our hybrid structure. We assume it 
to be uniform, which should be reasonable assumption as long as distance 
between the STM tip and the probed system was kept constant. 

\begin{figure}[!t]
\centering
\includegraphics[width=\linewidth]{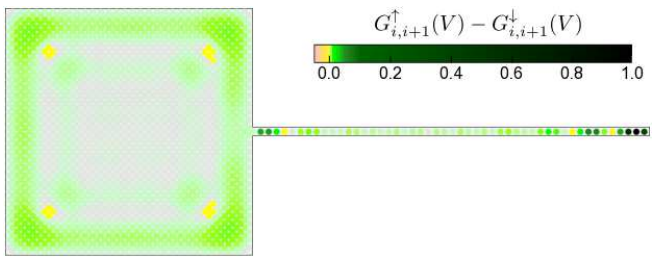}
\caption{
\label{sesar_plot}
Difference of the zero-bias tunnelling conductance between $\uparrow$
and $\downarrow$ charge carriers obtained in units of $2e^{2}/h$,
assuming both parts of our hybrid system to be in topologically 
non--trivial superconducting phase.}
\end{figure}

At low temperatures the conductance simplifies to  
\begin{eqnarray}
G_{i,i+1}^{\sigma}(V) \equiv 
\frac{d}{dV} I_{i,i+1}^{\sigma}(V) \simeq
\frac{2e^{2}}{h} T_{i,i+1}^{\sigma}(\omega = eV) .
\label{unconventional}
\end{eqnarray} 
Fig.~\ref{sesar_plot} shows a difference of the spin polarised 
conductance $G_{i,i+1}^{\uparrow}(V)-G_{i,i+1}^{\downarrow}(V)$
obtained at zero bias $V = 0$. Each point corresponds to the central 
place between the neighbouring sites $i$ and $i+1$ of the entire hybrid structure.
We clearly see that this polarised zero-bias conductance is
strongly enhanced near the localised Majorana mode at the end of nanowire 
and can also detect its delocalised partner, whose spectral weight is smeared along the boundaries of 2D-plaquette.
Additionally, it should be noted that the magnitude of spin polarised conductance is locally 100 times lower in plaquette than in nanowire.

Other  method for probing the delocalised mode could be based on measurements 
of the edge currents \cite{bjornson.pershoguba.15,pershoguba.bjornson.15,mohanta.kampf.18}, 
however we can hardly judge its practical feasability. Such local supercurrents
between the topologically trivial and non--trivial parts of the proximitized
systems have been also addressed in Ref.~\cite{cayao.blackschaffer.18} 
and several important aspects concerning the topological superconductivity  
of 2D systems have been discussed in 
Refs~\cite{sedlmayr.aguiarhualde.16,potter.lee.10}.

\section{Summary}
\label{sec.sum}

We have investigated quasiparticle spectra of the hybrid system, comprising 
the 1D-nanowire attached to the 2D-plaquette both proximitized to the  s-wave 
superconductor. Depending on the electron energies in these constituents 
(which should be tunable by external gate potentials), the spin-orbit 
interactions along with the magnetic field could induce the topologically 
non--trivial superconducting phase, either (i) only in the nanowire or 
(ii) in the entire setup. Selfconsistent numerical determination 
of the quasiparticle spectra has revealed, that under such circumstances 
the zero-energy Majorana quasiparticles would be (i) localised near 
the ends of 1D-nanowire, or (ii) one of them would leak into 
the plaquette region. For the latter case we have inspected 
the spatial profile of the delocalised Majorana mode and found 
its signatures distributed along boundaries of the 2D-plaquette, 
with some preference towards its corners.  
We have shown that, spatial profiles of both the localised and 
delocalised Majorana quasiparticles could be probed by the polarised 
scanning tunnelling measurements, where such quasiparticles 
would be detectable via the Andreev scattering mechanism. 

Here proposed hybrid structure (and similar ones) could open new 
perspectives for studying the topological superconducting phase 
in complex geometries of mixed dimensionality and might shed 
an insight into itinerancy of the emerging Majorana modes 
\cite{rachel.mascot.17,mascot.cocklin.18}. Furthermore, such delocalisation 
of the Majorana modes could be practically used for realisation 
of their braiding by attaching a few nanowires to the plaquette region.

After the initial version of this manuscript has been submitted 
for refereeing, we became aware of recent experimental evidence for
the delocalised Majorana modes in Fe islands deposited on Re 
superconducting substrate~\cite{wiesendanger_etal.18}. We have 
also learnt about the first experimental realisation of the mixed 
(localised and delocalised) Majorana modes  
using Co-Si magnetic clusters on surface of the inhomogeneous Pb 
superconductor~\cite{menard.17,PascalSimon_etal.18}. 
Recently, D.K.~Morr with coworkers~\cite{morr.17} 
have independently proposed very similar concept, emphasising that magnetic nanowires attached to magnetic island could be useful for manipulating the Majorana modes and could allow detection of the topological invariants for 2D-region. 
 

\begin{acknowledgments}
We thank N.~Sedlmayr for useful remarks.
A.P. is grateful to Laboratoire de Physique des Solides (CNRS, Universit\'{e} Paris-Sud) for hospitality during a part of the work on this project.
This projects is supported by the National Science Centre (NCN, Poland) under grants 
UMO-2017/27/B/ST3/01911 (A.K., T.D.), and UMO-2016/23/B/ST3/00647 (A.P.).
\end{acknowledgments}

\bibliography{biblio}

\end{document}